# BlueWalker 3 Redux

Anthony Mallama[1*], Richard E. Cole, Scott Harrington,
Aaron Worley, Jay Respler[1], Cees Bassa[1] and Scott Tilley[1]

2024 July 3

[1] IAU - Centre for the Protection of Dark and Quiet
Skies From Satellite Constellation Interference

*Contact: anthony.mallama@gmail.com

Abstract

The BlueWalker 3 satellite is now fainter than during the first months after deployment. The greatest improvement is that the average maximum luminosity near zenith has been reduced from magnitude 1.0 to 2.2. However, the spacecraft is still usually bright enough to interfere with astronomical research.

1. Introduction

The launch of BlueWalker 3 (BW3) in 2022 raised concerns in the astronomical community about its potential brightness. When the satellite deployed from a compact folded object into a giant 64 square meter flat panel in November of that year those concerns were realized. Early sightings reported that BW3 reached first magnitude when it was near the zenith (Mallama et al. 2022 and Nandakumar et al. 2023). Satellites brighter than magnitude 7 interfere with astronomical observing according to Tyson et al (2022) and BW3 exceeded that luminosity by more than 100 times.

Mallama et al. (2023a and 2023b) showed that the brightness of BW3 depends strongly on its orientation (attitude). The flat panel nominally faces zenith and nadir but sometimes it is tilted to increase sunlight on the solar power array situated on the upper side, as illustrated in Figure 1. At such times, the satellite can be dimmer than magnitude 7. Numerical modeling indicated that tilting was in the range of 13° to 16°.

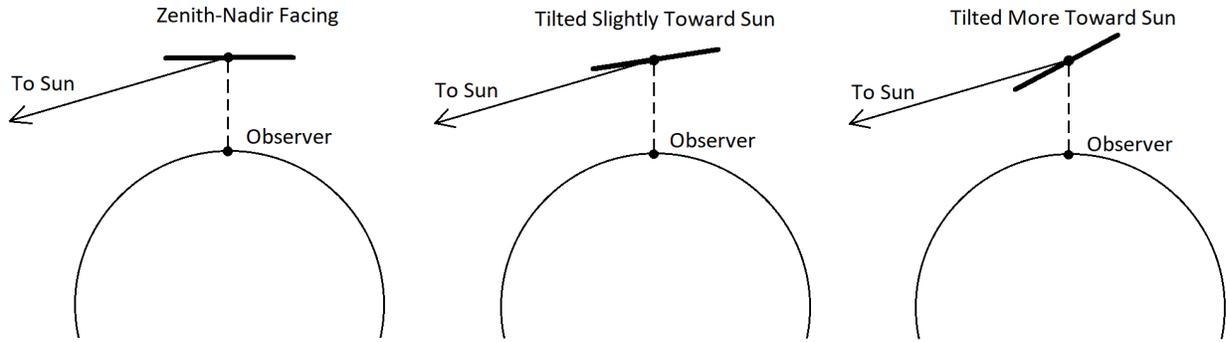

*Figure 1. The left diagram shows the nominal spacecraft attitude where the flat panel is zenith-and-nadir facing. There is no insolation on the solar array in that case. The middle illustrates an adjusted attitude where the zenith facing side of the panel is tilted slightly toward the Sun and the nadir side receives less insolation. In the right diagram, the panel is tilted further; no sunlight reaches the nadir side and the satellite is invisible. This Sun-satellite-observer geometry represents a satellite seen near zenith at the time boundary between astronomical twilight and darkness.*

We have continued monitoring the brightness of BW3 until the present time and the MMT9 robotic observatory (Karpov et al 2015 and Beskin et al 2017) has also recorded recent magnitudes. The light curve shown in Figure 2 suggests that the typical brightness of BW3 has declined over time. This implies that the spacecraft operator has increased the tilt.

Section 2 of this paper describes the observations. Section 3 examines brightness as a function of beta angle, and presents a light curve that factors in the effect of beta. Section 4 discusses the results and presents our conclusions.

2. Observations

The magnitudes analyzed in this study were obtained using electronic and visual observing methods. Electronic measurements were obtained at the MMT9 robotic observatory (Karpov et al. 2015 and Beskin et al. (2017). MMT9 consists of nine 71 mm diameter f/1.2 lenses and 2160 x 2560 sCMOS sensors. The magnitudes are within 0.1 of the V-band based on information in a private communication from S. Karpov as discussed by Mallama (2021).

Visual magnitudes were obtained by comparing spacecraft brightness to that of nearby reference stars. The angular proximity between satellites and stellar objects accounts for variations in sky transparency and sky brightness. The visual method of observing is described in more detail by Mallama (2022).

The apparent magnitudes were adjusted to a standard distance of 1000 km. Figure 2 reveals that BW3 has become fainter over time.

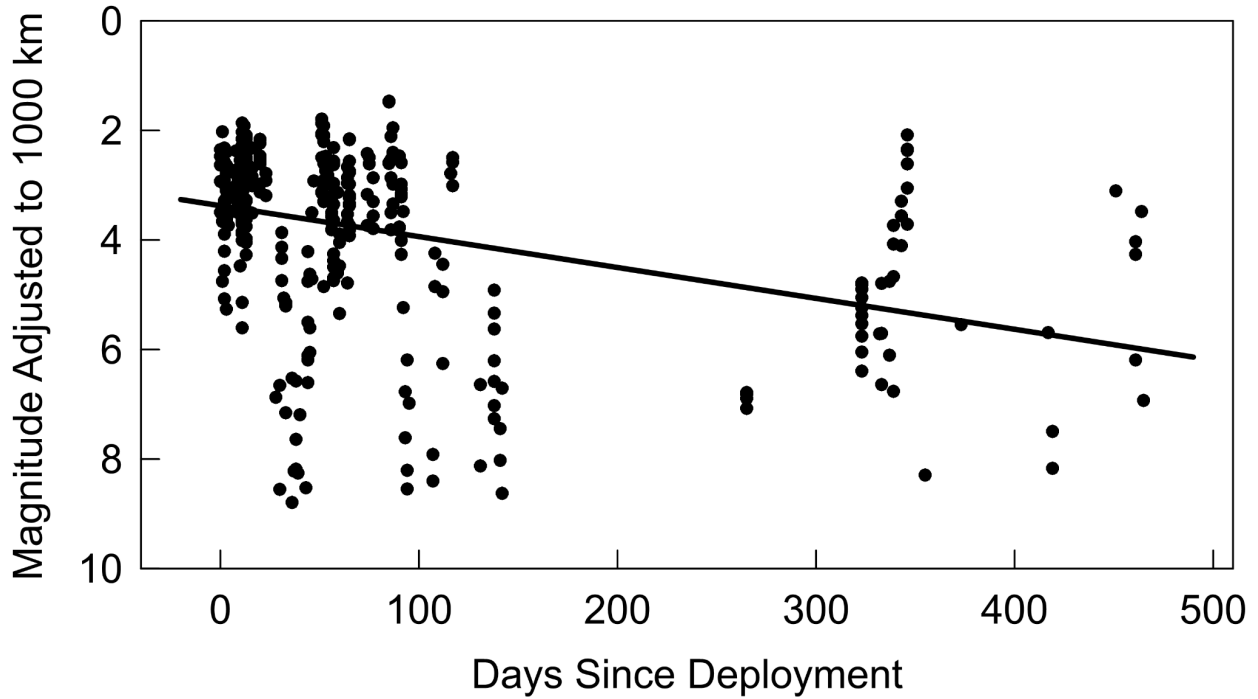

*Figure 2. Apparent magnitudes are adjusted to a standard distance of 1000 km. The linear fit indicates a downward trend in brightness. BW3 was launched as a small, compact object on 2022 September 11 and it deployed into a large flat-panel shape on 2022 November 11.*

3. Beta angle and brightness

The angle between the plane of a satellite orbit and the direction to the Sun is called *beta*. When the beta angle of BW3 is large, insolation on the solar power array situated on the upper side of the zenith-nadir pointing panel, is reduced. The satellite operator tilts the upper side toward the Sun in order to provide adequate power to the satellite at such times. The tilt reduces or eliminates sunlight on the lower side of the panel which faces observers on the ground as illustrated in Figure 1. The empirical relationship between the cosine of beta and distance-adjusted magnitude is shown in Figure 3.

The magnitudes plotted in Figure 2 suggest that a brightness change took place around 2023 February 19 which was 100 days after the satellite deployed into its large flat panel shape. The right-hand plot of Figure 3 shows that the function of brightness versus beta differs significantly before and after that date. The coefficients of the linear fits are listed in Table 1. The confidence lines on either side of the early and later fit lines indicate a significant difference at large values of cosine beta, where the satellite is most luminous.

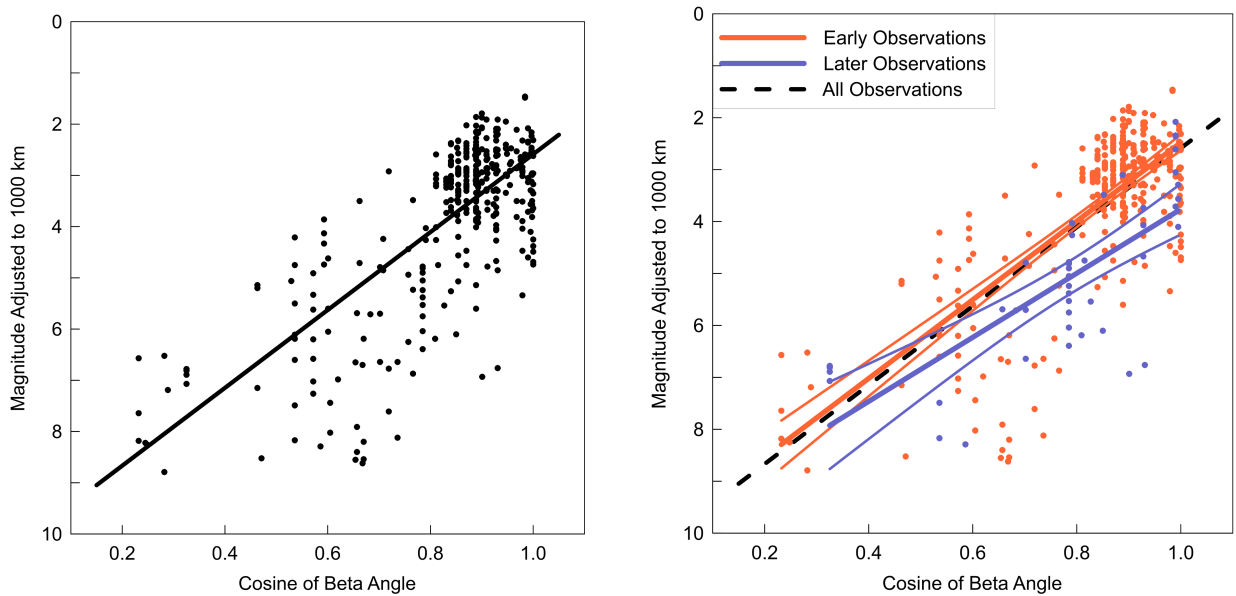

*Figure 3. The diagram at left plots all magnitudes as a function of cosine beta. On the right they are separated into those before and after 100 days past deployment. Thin lines are the confidence intervals.*

The residuals to the trend line for all magnitudes in Figure 3 are plotted against time in Figure 4. The brightness difference before and after 100 days past deployment is evident.

Table 1. Coefficients of magnitude versus cosine beta

|       | Intercept | Slope |
|-------|-----------|-------|
| Early | 10.05     | -7.57 |
| Later | 10.31     | -6.59 |
| All   | 10.30     | -7.72 |

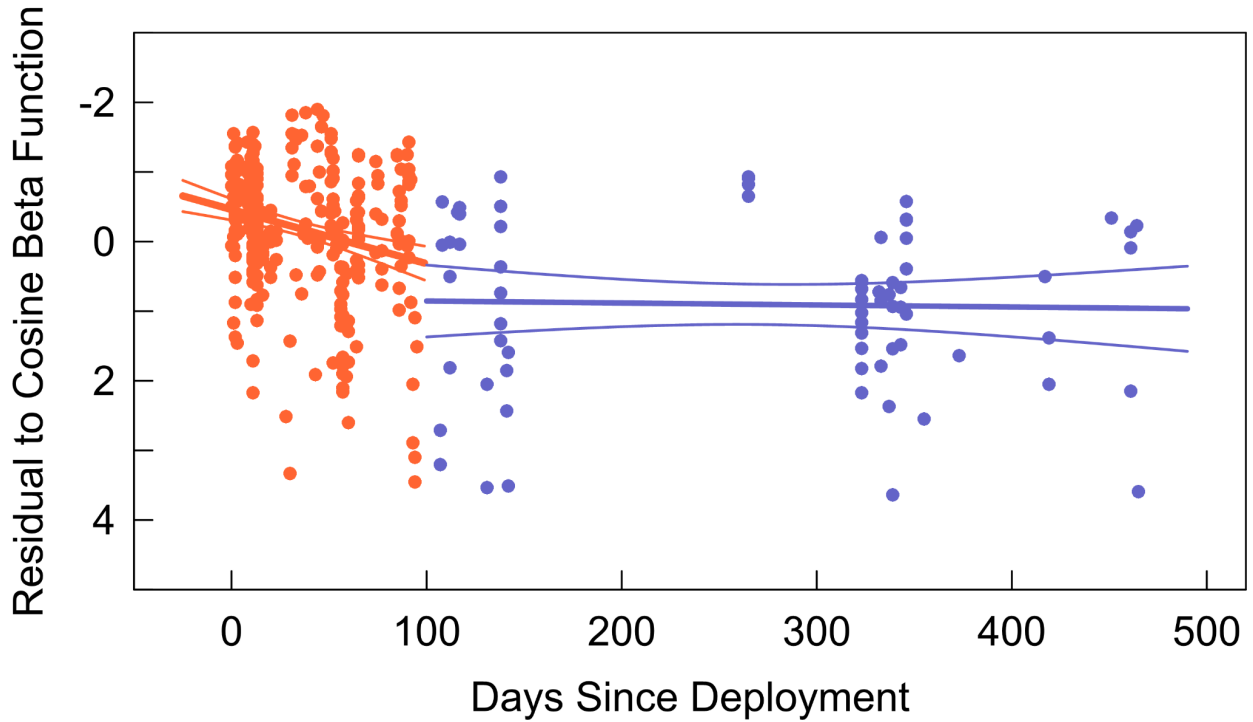

*Figure 4. Residuals to magnitude versus cosine beta for all observations indicate that later magnitudes are fainter.*

4. Discussion and Conclusions

BW3 is a concern for astronomers because its great brightness interferes with celestial observations. Satellites brighter than magnitude 7 seriously degrade images from wide field instruments such as the LSST as described by Tyson et al (2002). For casual sky watchers the limit is magnitude 6 because more luminous satellites are visible to the unaided eye.

In order to evaluate how often BW3 exceeds these brightness criteria, its beta angle was computed for every day over a period of one year. Then the corresponding 1000-km magnitude was computed based on the early and later fit coefficients listed in Table 1. Brightness computed from the early best fit exceeds the magnitude 6 and 7 limits 86% and 93% of the time, respectively. For the later best fit, these values reduce to 76% and 88%. At the zenith distance for BW3 of approximately 500 km, these percentages would be higher. The results indicate that tilting the satellite has reduced its impact on astronomy although it will still have a negative impact much of the time.

Most of the dimming has been achieved at values of cosine beta near unity where BW3 is brightest. The early observations indicate that the average magnitude of BW3 near zenith was 1.0. The newer observations reduce that computed brightness to magnitude 2.2.

The satellite operator, AST SpaceMobile issued a [press release](press release) on 2024 April 1 stating that 5 Block 1 BlueBird satellites will be shipped to the launch site between July and August of this year. BlueBirds are the follow-on to BW3 and they are the same size. AST also has secured a launch contract for Block 2 BlueBirds which will be more than 3 times as large as BW3.

We sent a message with a draft of this manuscript to AST as a courtesy and asked for details about their brightness mitigation plan. They forwarded the message to their public relations firm, AllisonWorldWide.com. The reply from the PR company contained only the following generic statement, "Thank you for the opportunity to review the manuscript in advance. AST SpaceMobile is committed to continuing to work with astronomers as the company advances its mission to democratize access to knowledge and information by providing space-based broadband to the world." They provided no details about brightness mitigation.

Acknowledgments
We thank the staff of the MMT9 robotic observatory for making their data available on-line. The Heavens-Above.com web-site was used to plan observations. Stellarium, Orbitron and QuickSat were used for data processing.

Mallama, A. 2021. Starlink satellite brightness– characterized from 100,000 visible light magnitudes. https://arxiv.org/abs/2111.09735.

Mallama, A., 2022. The method of visual satellite photometry. https://arxiv.org/abs/2208.07834.

Mallama, A., Cole, R.E., Harrington, S. and Maley, P.D. 2022. Visual magnitude of the BlueWalker 3 satellite. https://arxiv.org/abs/2211.09811.

Mallama, A., Cole, R.E. and Tilley, S. 2023a. The BlueWalker 3 satellite has faded. https://arxiv.org/abs/2301.01601.

Mallama, A., Cole, R.E., Tilley, S., Bassa, C., and Harrington, S. 2023b. BlueWalker 3 satellite brightness characterized and modeled. https://arxiv.org/abs/2305.00831.

Nandakumar,S. and 28 co-authors. 2023. The high optical brightness of the BlueWalker 3 satellite. Nature Astronomy **623,** 938-941. https://www.nature.com/articles/s41586-023-06672-7#:~:text=BlueWalker%203%20features%20a%2064.3%E2%80%89m%202%20phased-array%20antenna,of%20the%20brightest%20objects%20in%20the%20night%20sky.

Tyson, J. A. and 10 co-authors. 2020. Mitigation of LEO satellite brightness and trail effects on the Rubin Observatory LSST. Astron. J. 160, 226 and https://arxiv.org/abs/2006.12417.

6